\documentclass[preprint,showpacs,
preprintnumbers,amsmath,amssymb]{revtex4}
\usepackage{graphicx}
\usepackage{dcolumn}
\usepackage{bm}
\usepackage[usenames]{color}
\begin{document}
\title{Traveling kinks in cubic nonlinear Ginzburg-Landau equations}
\author{H. C.  Rosu}
\email{hcr@ipicyt.edu.mx}
\affiliation{IPICYT, Instituto Potosino de Investigacion Cientifica y Tecnologica,\\Apdo Postal 3-74 Tangamanga, 78231 San Luis Potos\'{\i}, S.L.P., Mexico.}
\author{O. Cornejo-P\'erez}
\affiliation{Facultad de Ingenier\'{\i}a, Universidad Aut\'onoma de Quer\'etaro, Centro Universitario Cerro de las Campanas,\\
76010 Santiago de Quer\'etaro, Mexico.}
\author{P. Ojeda-May}
\affiliation{Department of Chemistry, Indiana University--Purdue University, Indianapolis, USA.}
\date{\today}
\begin{abstract}
Nonlinear cubic Euler-Lagrange equations of motion in the traveling variable are usually derived from Ginzburg-Landau free energy functionals frequently encountered in several fields of physics. Many authors considered in the past damped versions of such equations with the damping term added by hand simulating the friction due to the environment. It is known that even in this damped case kink solutions can exist.
By means of a factorization method, we provide analytic formulas for several possible kink solutions of such equations of motion in the undriven and constant field driven cases, including the recently introduced Riccati parameter kinks which were not considered previously in such a context. The latter parameter controls the delay of the switching stage of the kinks. The delay is caused by antikink components that are introduced in the structure of the solution through this parameter.
\end{abstract}
\pacs{05.45.Yv, 75.60.Ch, 77.80.Fm\\
{\small arXiv:1107.4773 v4} $\qquad$ File: {\small susy-kinksF6} \hfill Phys. Rev. E 85, 037102 (2012).
}
\maketitle


Nonlinear field excitations occur in a rich variety of collective phenomena.
Perhaps the most prominent are the fields with cubic nonlinear equations of motion because they have effective potential energy with two minima of the type $B\psi^4 -A\psi^2$ and are commonly used as order parameters in the frame of the Ginzburg-Landau theory for the study of ferrodistortive domain walls \cite{collins}, in the analysis of the phase separation in binary mixtures \cite{van}, for diamagnetic (Condon) domains \cite{gordon}, and in more general situations such as domain walls in nonequilibrium systems \cite{hagberg}. A recently proposed discrete model for the curvature modes along protein backbone chains used by Chernodub {\em et al}. \cite{chern} to explain protein folding belongs to the same approach. The dynamics of all these systems can be treated variationally as a relaxation process toward one of the stationary states in the potential wells. This relaxation is governed by equations of the type $\partial\psi/\partial t\propto -\delta {\cal F}/\delta \psi$ where ${\cal F}$ is the free energy of these systems.

In this work, we focus on the solutions of the following equations of motion in the traveling coordinate $\xi=x-vt$  and with rescaled coefficients
\begin{equation}\label{ec1}
  \psi''  + \rho \psi' - B_1\psi^3 + A_1\psi=0  
\end{equation}
and
\begin{equation}\label{ec2}
  \psi''  + \rho \psi' - B_1\psi^3 + A_1\psi+\gamma_1\eta=0  
\end{equation}
if a constant external field $\eta$ multiplied by its scaled coupling constant $\gamma_1$ is added.
Except for the friction term, these equations are Euler-Lagrange equations of motions corresponding to Ginzburg-Landau functionals.
The derivation of such equations can be found for example in Ref. \cite{collins} in the context of ferrodistortive domain walls, while recently, Mavromatos \cite{mavro} discussed a well-known counterpart of (\ref{ec2}) in the case of microtubules where the friction is attributed to the so-called ordered water molecules. We make clear that the friction coefficient $\rho$, although constant, depends on the (constant) velocity of the frame. Both relativistic and nonrelativistic dependencies can be encountered in the literature. For example, in the case of ferrodistortive materials $\rho\propto v(c_{0}^{2}-v^2)^{-1/2}$, where $c_{0}$ is the limiting velocity of the system, whereas $\rho\propto v(D-mv^2/2)^{-1}$, where $D$ is a diffusion coefficient and $m$ is the inertia parameter, in the case of fast, nonoverdamped motion of spin domain walls in Ising ferromagnetics \cite{gvw}.
For equations without the friction term, kink solutions have been well known for at least 40 years \cite{montroll}. In fact, Montroll \cite{montroll} mentions that Fisher already investigated numerically solutions of the Fisher equation with a first derivative term. The existence of kink solutions in the presence of friction terms has been well settled since the works of Lal \cite{Lal}, Geicke \cite{g85}, and Kashcheev \cite{K88}. The first goal of this work is to show that for the Eqs.~(\ref{ec1}) and (\ref{ec2}) various kink solutions can be easily obtained by a factorization technique that we introduced previously \cite{rosu1} for equations of the form ($D_s={d}/{d s}$)
\begin{equation}\label{n1}
D^{2}_{s} \psi+\rho D_s \psi  +F({\psi})=0,
\end{equation}
where  $F({\psi})$ is a polynomial function, which in the case (\ref{ec1}) and (\ref{ec2}) is a cubic polynomial.
Equation (\ref{n1}) can be  factorized as follows:
\begin{equation}
[D_s-f_{2}({\psi})][D_s-f_{1}({\psi})]\psi(s)=0~.\label{n2}
\end{equation}
Expanding (\ref{n2}), one can use the following grouping of terms \cite{rosu1}:
\begin{equation}\label{n4}
D^{2}_{s} \psi -\left(f_{1}+f_{2}+
\frac{df_{1}}{d\psi} \psi \right)D_s \psi+f_{1}f_{2}\psi=0~,
\end{equation}
and comparing Eq.~(\ref{n1}) with Eq.~(\ref{n4}), we get the conditions
\begin{eqnarray}
&&f_{1}(\psi)\, f_{2}(\psi)=\frac{F(\psi)}{\psi},\label{n5}\\
&&f_{2}(\psi)+ \frac{d(f_{1}(\psi)\, \psi)}{d\psi}= -\rho.\label{n6}
\end{eqnarray}
Any factorization like (\ref{n2}) of a scalar equation with polynomial nonlinearities of the form given in Eq. (\ref{n1}) allows us to find a compatible first order nonlinear differential equation,
\begin{equation}\label{n7}
[D_s-f_{1}(\psi)]\psi=D_s \psi-f_{1}(\psi)\psi=0~,
\end{equation}
whose solution provides a particular solution of (\ref{n1}).
In other words, if by some means we are able to find a couple of functions $f_{1}(\psi)$ and $f_{2}(\psi)$ such that they factorize Eq.~(\ref{n1}) in the form (\ref{n2}), solving Eq.~(\ref{n7}) allows to get particular solutions of (\ref{n1}). The advantage of this factorization is that the
two unknown functions $f_{1}(\psi)$ and $f_{2}(\psi)$ can be found easily by factoring the polynomial expression  (\ref{n5})
in terms of  linear combinations in rational powers of $\psi$. This technique is used in the following 
to find kink solutions in the undriven case and the constant field driven case. 
We also discuss the kinks based on the general Riccati solution, which depend on a control parameter of the switching features. The latter kinks, which we call Riccati parameter kinks, have not been discussed previously in the Ginzburg-Landau framework and drawing the attention to them is another important motivation for this work.


We first consider the case of zero external field. Montroll showed that Eq.~(\ref{ec1}) has a unique bounded (kinklike) solution of the form \cite{montroll}

\begin{equation}\label{Mont1}
\psi_M(\xi) = a + \frac{\sqrt{2}\alpha}{1+\exp (\alpha \xi)}~,
\end{equation}
where $\alpha=(b-a)/\sqrt{2}$ and the parameters $a$ and $b$ are two of the solutions of the cubic equation
\begin{equation}
(\psi - a)(\psi - b)(\psi - d) =  \psi^3-\psi~. 
\end{equation}
Taking $a=0$, $b=1$ and $d=-1$ gives $\alpha=\frac{1}{\sqrt{2}}$. The Montroll kink $\psi _M$ can be easily derived through the factorization procedure just described. Indeed, Eq.~(\ref{ec1}) can be factorized in the following two forms ($D_\xi=\frac{d}{d\xi}$)
\begin{equation}\label{ec2bis}
\left[ D_\xi \pm 2 ^{\frac{1}{2}}(\sqrt{A_1}+\sqrt{B_1}\psi)\right]\left[ D_\xi \pm
2 ^{-\frac{1}{2}}(\sqrt{A_1}-\sqrt{B_1}\psi)\right]\psi=0
\end{equation}
and
\begin{equation}
\left[ D_\xi \pm 2 ^{\frac{1}{2}}(\sqrt{A_1}-\sqrt{B_1}\psi)\right]\left[ D_\xi \pm
2 ^{-\frac{1}{2}}(\sqrt{A_1}+\sqrt{B_1}\psi)\right]\psi=0.\label{ec3}
\end{equation}
However, factorizations (\ref{ec2bis}) and (\ref{ec3}) are only
possible for $\rho_{\pm}=\pm\frac{3\sqrt{2}}{2}\sqrt{A_1}$ as obtained from (\ref{n6}).

Equation (\ref{ec2bis}) is compatible with the Riccati equations
\begin{equation}
\psi^{\prime} \pm 2 ^{-\frac{1}{2}}(\sqrt{A_1}\psi-\sqrt{B_1}\psi^2)=0,\label{ec4}
\end{equation}
whose (particular) solutions are
\begin{equation}
\psi_{1,2}=\frac{\sqrt{A_1}}{\sqrt{B_1}+\textrm{e}^{\pm\sqrt{A_1}(\xi-\xi _0)/\sqrt{2}}}
\equiv \frac{\sqrt{A_1}}{\sqrt{B_1}+k_1\textrm{e}^{\pm\sqrt{A_1}\xi/\sqrt{2}}}~, \quad k_1 =e^{\mp \sqrt{A_1}\xi_0/\sqrt{2}}~.\label{ec5}
\end{equation}

On the other hand, the compatible Riccati equations for (\ref{ec3}) are
\begin{equation}
\psi^{\prime} \pm 2 ^{-\frac{1}{2}}(\sqrt{A_1}\psi+\sqrt{B_1}\psi^2)=0~,\label{ec6}
\end{equation}
with the particular solutions
\begin{equation}
\psi_{3,4}=\frac{\sqrt{A_1}} 
{-\sqrt{B_1}+\textrm{e}^{\mp \sqrt{A_1}(\xi-\xi_0)/\sqrt{2}}}
\equiv \frac{\sqrt{A_1}} 
{-\sqrt{B_1}+\frac{1}{k_1}\textrm{e}^{\mp \sqrt{A_1}\xi/\sqrt{2}}}~.\label{ec7}
\end{equation}
All these solutions are similar to those given by Geicke \cite{g85} for this case and one can notice that $\psi _1$ corresponds to the Montroll kink of parameters $(0,1,-1)$ when $A_1=1$ and $B_1=1$.

\medskip


Moving now to the more complicated case given by Eq.~(\ref{ec2}), let $\gamma_1\eta=A_1\epsilon -B_1\epsilon^3$ and $\varphi=\psi +\epsilon$ (see also \cite{fp06}). Then, we get
\begin{equation}\label{a2-1}
\varphi ''+\rho\varphi '-\varphi\left(B_1\varphi ^2-3B_1\epsilon\varphi +(3B_1\epsilon^2-A_1)\right)=0~.
\end{equation}

{\em Case I}. The factorization of (\ref{a2-1}) can be achieved with
$f_1=a_1i\left[\sqrt{B_1}\varphi-r_+(\epsilon)
\right]$ and $f_2=a_{1}^{-1}i\left[\sqrt{B_1}\varphi-r_-(\epsilon)
\right]$, where
\begin{equation}\label{r+-}
r_{\pm}(\epsilon)=\frac{3\sqrt{B_1}\epsilon\pm(4A_1-3B_1\epsilon^2)^{\frac{1}{2}}}{2}\equiv\frac{3\sqrt{B_1}\epsilon\pm \sqrt{\Delta_{\epsilon}}}{2}~.
\end{equation}
From the second factorization condition (\ref{n6}), 
we get
\begin{equation}\label{rho-}
a_1=\pm 2 ^{-\frac{1}{2}}i \longrightarrow \rho _{\pm}^{(-)} =\pm 2 ^{-\frac{1}{2}}\left(r_{-}(\epsilon)-\sqrt{\Delta_{\epsilon}}\right)~.
\end{equation}
To have real values of the parameter $\rho _{\pm}^{(-)}$, one requires $\epsilon^2\leq 4A_1/3B_1$, and one gets a positive-valued friction parameter if $\epsilon\in \left(\sqrt{\frac{A_1}{B_1}},\frac{2}{\sqrt{3}}\sqrt{\frac{A_1}{B_1}}\right]$ for the positive front sign of $\rho$ and $\epsilon\in \left[-\frac{2}{\sqrt{3}}\sqrt{\frac{A_1}{B_1}},\sqrt{\frac{A_1}{B_1}}\right)$ for the negative front sign of $\rho$.
We are led to the following Riccati equations:
\begin{equation}\label{R1d}
\varphi _\xi \pm \sqrt{\frac{B_1}{2}}\varphi ^2\mp \sqrt{\frac{B_1}{2}}r_{+}(\epsilon)\varphi=0~,
\end{equation}
having as particular solutions
\begin{equation}\label{extfield1}
\varphi _{1}^{\pm}=\frac{1}{\sqrt{B_1}}\frac{2r_{+}(\epsilon)}{2+e^{\mp \alpha _1(\xi-\xi_0)}}~, \qquad \alpha _1=\frac{r_+(\epsilon)}{\sqrt{2}}~,
\end{equation}
where $\xi _0$ is a constant of integration. The last step is to go back to the $\psi$ solution, $\psi_{1}^{\pm}=\varphi _{1}^{\pm}-\epsilon$.
Solutions (\ref{extfield1}) were first found by Geicke through an ansatz method \cite{g85}.

{\em Case II}. The factorization of (\ref{a2-1}) can be also achieved with  $f'_1=a_{1}^{2}f_2$ and $f'_2=a_{1}^{-2}f_1$.
Using condition (\ref{n6}), one gets
\begin{equation}\label{rho+}
a_1=\pm 2 ^{-\frac{1}{2}}i \longrightarrow\rho _{\pm}^{(+)}=\pm 2 ^{-\frac{1}{2}}\left[r_+(\epsilon)+\sqrt{\Delta_\epsilon}\right]~.
\end{equation}
Real values of the parameter $\rho _{\pm}^{(+)}$ are again obtained for $\epsilon^2\leq \frac{4A_1}{3B_1}$, which implies a positive-valued friction parameter if $\epsilon\in \left(-\sqrt{\frac{A_1}{B_1}},\frac{2}{\sqrt{3}}\sqrt{\frac{A_1}{B_1}}\right]$ for the positive front sign of $\rho$ and if $\epsilon\in \left[-\frac{2}{\sqrt{3}}\sqrt{\frac{A_1}{B_1}},-\sqrt{\frac{A_1}{B_1}}\right)$ for the negative front sign of $\rho$.


This factorization implies Riccati equations of the form:
\begin{equation}\label{R2d}
\varphi _\xi \pm 2 ^{-\frac{1}{2}}\varphi ^2\mp 2 ^{-\frac{1}{2}}r_{-}(\epsilon)\varphi=0~,
\end{equation}
whose particular solutions are
\begin{equation}\label{3-last one}
\varphi _{2}^{\pm}=\frac{1}{\sqrt{B_1}}\frac{2r_{-}(\epsilon)}{2+e^{\mp \alpha _2(\xi-\xi_0)}}~, \qquad \alpha _2=\frac{r_-(\epsilon)}{\sqrt{2}}~.
\end{equation}


The factorization method allows even more general solutions, the so-called Riccati parameter solutions introduced by Reyes and Rosu \cite{rr08} which are based on the general Riccati solution. Indeed, all Riccati equations in this Brief Report are of constant coefficients,
say $y'-c_1y^2-c_2y =0$, and if a particular solution $y_1$ is known then the general solution depending on a free parameter denoted by $\lambda$ can be written as
\begin{equation}\label{lambda rr1}
y_{\lambda, c_1,c_2}=y_1+\frac{e^{I_1}}{\lambda-c_1I_2}~,
\end{equation}
where $I_1=\int_{\xi_0}^{\xi}\left(2c_1y_1+c_2\right)dx$ and $I_2=\int _{\xi_0}^{\xi}e^{I_1(x)}dx$.
For the nonlinear equations of motion discussed here the formulas for the kinks given by Eq.~(\ref{lambda rr1}) are as follows.

(a) Zero field (kinks having mixtures of rising and decaying exponentials, both of width $\alpha^{-1} =\sqrt{2/A_1}$): 

 \begin{equation}\label{lambda 01}
 \psi ^{+}_{1,\lambda}=\frac{\sqrt{A_1}}{\mp\sqrt{B_1}+e^{\alpha(\xi - \xi _0)}}\left[1+\frac{1}{\lambda \sqrt{A_1}(1\mp\sqrt{B_1}e^{-\alpha(\xi - \xi _0)})\mp\sqrt{B_1}e^{-\alpha(\xi - \xi _0)}}\right]~,
 \end{equation}
 where the minus sign corresponds to the first factorization and the plus sign corresponds to the second one.
\begin{equation}\label{lambda 02}
 \psi ^{-}_{1,\lambda}=\frac{\sqrt{A_1}}{\mp\sqrt{B_1}+e^{-\alpha(\xi - \xi _0)}}\left[1+\frac{1}{\lambda \sqrt{A_1}(1\mp\sqrt{B_1}
 e^{\alpha(\xi - \xi _0)})-1}\right]~,
\end{equation}
with the same rule of signs.

(b) Driving constant field.

{\em Case I}. (kinks having mixtures of rising and decaying exponentials both of width $\alpha_{1}^{-1} =\sqrt{2}/r_{+}(\epsilon)$):\\
\begin{equation}\label{driven C1}
\varphi_{\lambda}^{+}=\frac{2r_{+}(\epsilon)}{2+e^{- \alpha _1(\xi-\xi_0)}}\left[\frac{1}{\sqrt{B_1}}+\frac{1}{2\lambda r_{+}(\epsilon)\left[1+2e^{+ \alpha _1(\xi-\xi_0)}\right]-\sqrt{B_1}}\right]
\end{equation}
and
\begin{equation}\label{driven C2}
\varphi_{\lambda}^{-}=\frac{2r_{+}(\epsilon)}{2+e^{+ \alpha _1(\xi-\xi_0)}}\left[\frac{1}{\sqrt{B_1}}+\frac{1}{2\lambda r_{+}(\epsilon)\left[1+2e^{- \alpha _1(\xi-\xi_0)}\right]+2\sqrt{B_1}e^{- \alpha _1(\xi-\xi_0)}}\right]~.
\end{equation}
These kinks do not have singularities  if $\lambda \notin\left(0, \frac{\sqrt{B_1}}{2r_+}\right]$ and
$\lambda \notin\left[-\frac{\sqrt{B_1}}{2r_+}, 0\right)$, respectively.
The parametric solutions $\psi_{\lambda}^{\pm}$ are obtained immediately by downshifting the above solutions by $\epsilon$. Plots of solutions $\psi_{\lambda}^{+}$ and $\psi_{\lambda}^{-}$ are displayed in Figs. 1 and 2, respectively.

{\em Case II}. (kinks with mixtures of rising and decaying exponentials both of width $\alpha_{2}^{-1} =\sqrt{2}/r_{-}(\epsilon)$):\\ The parametric solutions turn out to be
\begin{equation}\label{driven C2-1}
\varphi_{\lambda}^{+}=\frac{2r_{-}(\epsilon)}{2+e^{-\alpha _2(\xi-\xi_0)}}\left[\frac{1}{\sqrt{B_1}}+\frac{1}{2\lambda r_{-}(\epsilon)\left[1+2e^{\alpha _2(\xi-\xi_0)}\right]-\sqrt{B_1}}\right]
\end{equation}
and
\begin{equation}\label{driven C2-2}
\varphi_{\lambda}^{-}=\frac{2r_{-}(\epsilon)}{2+e^{+ \alpha _2(\xi-\xi_0)}}\left[\frac{1}{\sqrt{B_1}}+\frac{1}{2\lambda r_{-}(\epsilon)\left[1+2e^{-\alpha _2(\xi-\xi_0)}\right]+2\sqrt{B_1}e^{-\alpha _2(\xi-\xi_0)}}\right]~.
\end{equation}
Choosing $\lambda \notin \left(0, \frac{\sqrt{B_1}}{2r_-}\right]$, the $\varphi_{\lambda}^{+}$ kink does not have any singularities, while in the case of $\varphi_{\lambda}^{-}$ the forbidden interval for $\lambda$ is $\left[ -\frac{\sqrt{B_1}}{2r_-}, 0\right)$.
Again, the corresponding $\psi$ parametric solutions are obtained by downshifting by $\epsilon$.
Plots of the solutions $\psi_{\lambda}^{+}$ and $\psi_{\lambda}^{-}$ in this case are displayed in Figs. 3 and 4, respectively.

   \begin{figure}[x!]
     \centering
     \includegraphics[width=7 cm, height=7 cm]{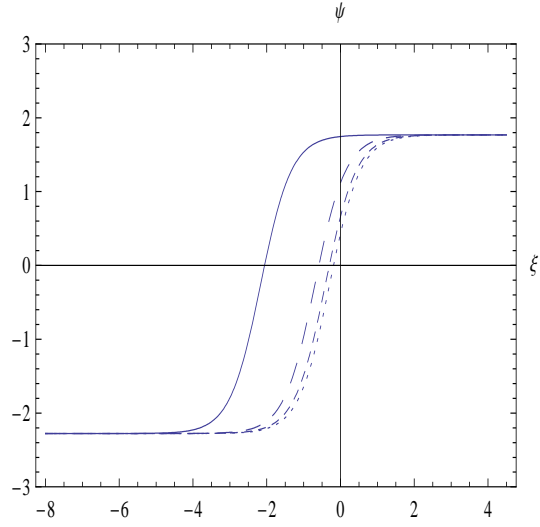}
     \caption{
     {\small Case I. Plot of $\psi^{+}_{\lambda} (\xi)$ for $\lambda=0.125$ (solid line), 0.2 (long-dashed line), 0.5 (dashed line) and 10 (dotted line). $A_1=3$, $B_1=0.7$, $\rho=0.90326$ ($\epsilon=2.2772$), and $\xi_0=0$.
     }}
     \label{tk1}
   \end{figure}

   \begin{figure}[x!]
     \centering
     \includegraphics[width=7 cm, height=7 cm]{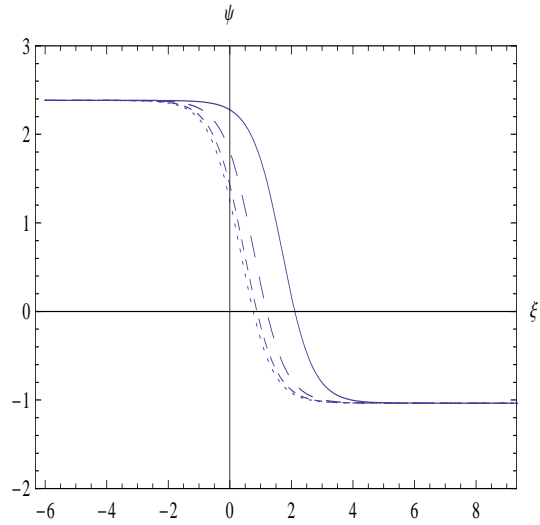}
     \caption{
     {\small Case I. Plot of $\psi^{-}_{\lambda} (\xi)$ for $\lambda=0.01$ (solid line), 0.1 (long-dashed line), 0.5 (dashed line) and 10 (dotted line). $A_1=3$, $B_1=0.7$, $\rho=2.39335$ ($\epsilon=1.0351$), and $\xi_0=0$.
     }}
     \label{tk2}
   \end{figure}

   \begin{figure}[x!]
     \centering
     \includegraphics[width=7 cm, height=7 cm]{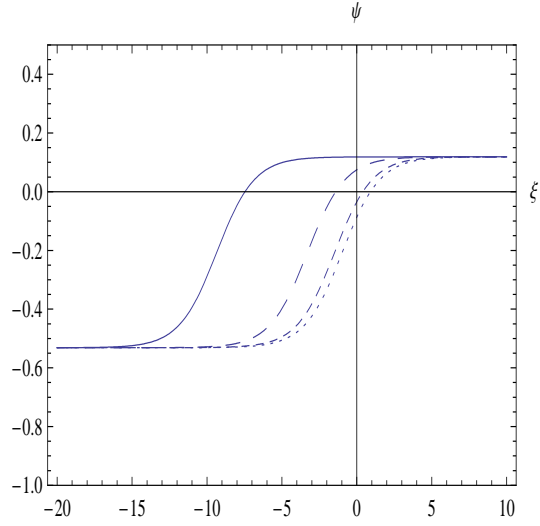}
     \caption{
     {\small Case II. Plot of $\psi^{+}_{\lambda} (\xi)$ for $\lambda=0.77$ (solid line), 0.9 (long-dashed line), 2 (dashed line) and 10 (dotted line). $A_1=0.7$, $B_1=3$, $\rho=1.51635$ ($\epsilon=0.5313$), and $\xi_0=0$.}}
     \label{tk3}
   \end{figure}

   \begin{figure}[x!]
     \centering
     \includegraphics[width=7 cm, height=7 cm]{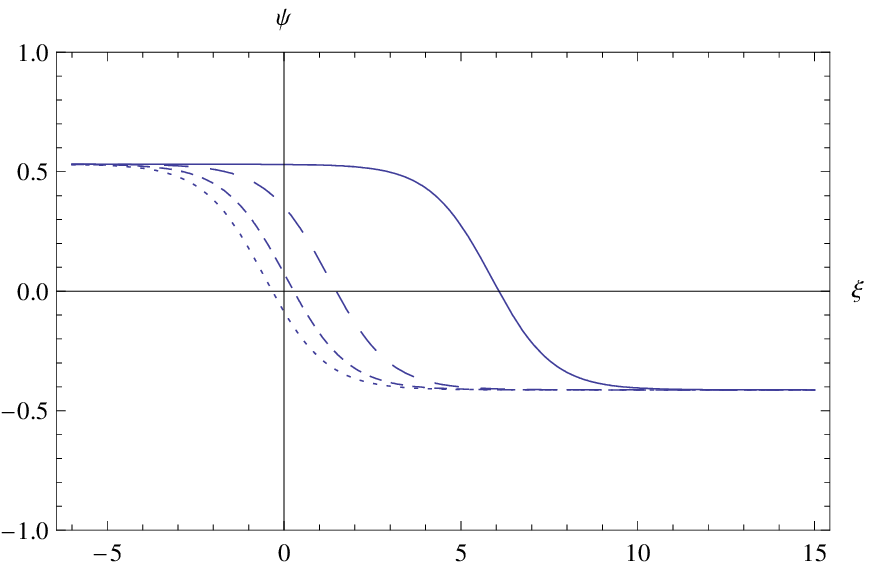}
     \caption{
     {\small Case II. Plot of $\psi^{-}_{\lambda} (\xi)$ for $\lambda=0.53$ (solid line), 0.6 (long-dashed line), 1 (dashed line) and 10 (dotted line). $A_1=0.7$, $B_1=3$, $\rho=0.435766$ ($\epsilon=-0.5313$), and $\xi_0=0$.
     }}
     \label{tk4}
   \end{figure}


We now briefly comment on the stability of the Riccati parameter kinks. The stability analysis depends on the form of the velocity dependence of the friction-like parameter $\rho$. If we take this dependence as in the paper of Collins {\em et al}. \cite{collins}, one can follow step by step the stability procedure presented therein. This is because at the first step of the stability analysis, that of writing a perturbed kink solution $y(\zeta, t)=y(\zeta; \lambda)+\delta y(\zeta,t)$ one notices that the Riccati kink $y(\zeta; \lambda)$ is a solution of the same equation as the common kink $y(\zeta; \lambda=\infty)$ and therefore the linearization leads to the same eigenvalue problem. For a velocity dependence of $\rho$ corresponding to a driven, damped, nonlinear Klein-Gordon type equation, the stability analysis is somewhat more complicated but has been sketched in the important paper of B\"uttiker and Thomas \cite{bt88}.

In summary, using the factorization method introduced in \cite{rosu1}, we have obtained the analytic forms of various kink solutions of the nonlinear cubic Euler-Lagrange equations of the damped type in the traveling variable. The results presented here can be directly applied to the Condon domains if we make the following identification of our parameters with the parameters given by Gordon {\em et al}. \cite{gordon}: $\rho=v/K\Gamma, A_1=A/K, B_1=B/K, \gamma_1=a/k$; in the case of Collins {\em et al}. \cite{collins} one should take $A_1=B_1=1$. Examining the formulas for the Riccati parameter kinks, one can easily infer that the parameter $\lambda$ occurs as a control parameter of the initiation of the switching stage \cite{rr08}. Indeed, $\lambda$ is associated with the exponentials of opposite exponent in the denominators with respect to the exponential of the particular Riccati solution and this is what generates the delay. Interestingly,
we notice that although the switching is more delayed when $\lambda$ increases, this is so only at relatively low values of the parameter, while at higher values of $\lambda$ the delay saturates. Since switching is related to microscopic restructuring of the kink (mesoscopic domain), one may think that the $\lambda$ parameter can characterize the dependence of the switching delay on the rate at which an applied field is ramped up or down.
Finally, all the kinks discussed here occur in conditions of environmental friction, which is not easy to define microscopically \cite{mavro}. If the frictional effects are considered as first derivative terms in cubic nonlinear equations of motion as done here, then the kinks discussed in this work occur only for very particular values of the friction coefficient given by $\rho_{\pm}$, $\rho_{\pm}^{(-)}$, and $\rho_{\pm}^{(+)}$.


{\small

\begin{thebibliography}{999}

\bibitem{collins} M. Collins, A. Blumen, J.F. Currie, J. Ross,
{\em Dynamics of domain walls in ferrodistortive materials. I. Theory},
Phys. Rev. B {\bf 19}, 3630-3644 (1979).

\bibitem{van}  Y. Huo, X. Jiang, H. Zhang, Y. Yang,
{\em Hydrodynamic effects on phase separation of binary mixtures with reversible chemical reaction},
 J. Chem. Phys. {\bf 118}, 9830-9837 (2003). 

\bibitem{gordon} A. Gordon, N. Logoboy, W. Joss,
{\em Size-dependent effects on the magnetization dynamics of Condon domains},
Phys. Rev. B {\bf 69}, 174417 (2004).

\bibitem{hagberg} A. Hagberg, E. Meron,
{\em Domain walls in nonequilibrium systems and the emergence of persistent patterns},
Phys. Rev. E {\bf 48}, 705-708 (1993).

\bibitem{chern} M. Chernodub, S. Hu, A. Niemi,
{\em Topological solitons and folded proteins},
Phys. Rev. E {\bf 82}, 011916 (2010).

\bibitem{mavro} N.E. Mavromatos,
{\em Quantum mechanical aspects of cell microtubules: science fiction or realistic possibility ?},
J. Phys. Conf. Series {\bf 306}, 012008 (2011).

\bibitem{gvw} A. Gordon, I.D. Vagner, P. Wyder, {\em On some pecularities in the dynamics of magnetic domains}, Solid St. Commun. {\bf 87}, 1155-1158 (1993).

\bibitem{montroll} E.W. Montroll,
{\em Nonlinear rate processes, especially those involving competitive processes},
in {\em Statistical Mechanics}, edited by S.A. Rice, K.F. Freed, and J.C. Light (University of Chicago Press, Chicago, 1972), pp. 69-91, especially pp. 84-87.

\bibitem{Lal} P. Lal,
{\em Kink solitons and friction},
Phys. Lett. A {\bf 111}, 389-390 (1985).

\bibitem{g85} J. Geicke,
{\em Travelling wave solutions to the perturbed $\phi^4$ equation},
Phys. Lett. A {\bf 111}, 10-14 (1985).

\bibitem{K88} V.N. Kashcheev,
{\em Kinks in systems with cubic and quartic anharmonicity},
Theor. Math. Phys. {\bf 74}, 43-48 (1988).

\bibitem{rosu1} H.C. Rosu, O. Cornejo-P\'erez,
{\em Supersymmetric pairing of kinks for polynomial nonlinearities},
Phys. Rev E {\bf 71}, 046607 (2005). O. Cornejo-P\'erez, H.C. Rosu,
{\em Nonlinear second order ODEs: Factorizations and particular solutions},
Prog. Theor. Phys. {\bf 114}, 533-538 (2005).


\bibitem{fp06} O. Cornejo-P\'erez, J. Negro, L.M. Nieto, H.C. Rosu,
{\em Traveling-wave solutions for Korteweg-deVries-Burgers equations through factorizations},
Found. Phys. {\bf 36}, 1587-1599 (2006).


\bibitem{rr08} M.A. Reyes, H.C. Rosu,
{\em Riccati-parameter solutions of nonlinear second order ODEs},
J. Phys. A {\bf 41}, 285206 (2008); see also W. Alka, A. Goyal, C.N. Kumar,
                     {\em Nonlinear dynamics of DNA-Riccati generalized solitary wave solutions},
                     Phys. Lett. A {\bf 375}, 480-483 (2011).

 \bibitem{bt88} M. B\"uttiker, H. Thomas,
 {\em Propagation and stability of kinks in driven and damped nonlinear Klein-Gordon chains},
Phys. Rev. A {\bf 37}, 235-246 (1988).


\end{thebibliography}

}
\end{document}